\documentclass[%
 aps,
 amsmath,amssymb,
 aps,
]{revtex4}
\usepackage{epsfig, color}

\usepackage{graphicx}
\graphicspath{%
    {converted_graphics/}
    {/}
}
\usepackage{amsfonts,amssymb,
amsthm}
\usepackage{amsmath}
\usepackage{graphicx}
\usepackage{enumerate}
\usepackage{subfigure}
\usepackage{mathtools}
\usepackage{amscd}
\usepackage[toc]{appendix}
\usepackage{afterpage}
\usepackage{bm}
\usepackage{epstopdf}
\usepackage{mathrsfs}
\newcommand{\eproof}{\hfill{\vrule height5pt width5pt depth0pt}\medskip}

\renewcommand{\proof}{{\noindent {\em Proof}.\;}}

\newtheorem{theorem}{Theorem}[section]

\begin{document}

\title[KdV-Burgers and Gardner]
{Dissipative periodic and chaotic  patterns to the \\ KdV--Burgers and Gardner equations}

\author{Stefan C. Mancas}
\email{mancass@erau.edu}
\affiliation{Department of Mathematics, Embry-Riddle Aeronautical University\\
Daytona Beach, FL 32114-3900, USA}

\author{Ronald Adams}
\email{adamsr25@erau.edu}
\affiliation{Department of Mathematics, Embry-Riddle Aeronautical University\\ Daytona Beach, FL 32114-3900, USA}

\begin{abstract}
We investigate the KdV-Burgers and Gardner equations  with  dissipation and external perturbation terms by the approach of dynamical systems and Shil'nikov's analysis.  The stability of the equilibrium point  is considered, and Hopf bifurcations are investigated after a certain scaling that reduces the parameter space of a  three-mode dynamical system which now depends only on two parameters.  The Hopf curve divides  the two-dimensional space into two regions. On the left region the equilibrium  point is stable leading to dissapative periodic orbits. While changing the bifurcation parameter given by the velocity of the traveling waves,   the equilibrium point  becomes unstable and a unique stable limit cycle bifurcates from the origin. This limit cycle is the result of a supercritical Hopf bifurcation which is proved using the Lyapunov coefficient together with the Routh-Hurwitz  criterion.  On the right side of the Hopf curve, in the case of the  KdV-Burgers, we find homoclinic chaos by using  Shil'nikov's theorem which requires the construction of a homoclinic orbit, while for the Gardner equation the supercritical Hopf bifurcation leads only to a stable periodic orbit.

\end{abstract}
\maketitle

\section{Introduction}
We  consider the nonlinear evolution equation
\begin{equation} \label{eq1}
\frac{\partial u}{\partial t}+ \alpha u^n \frac{\partial u}{\partial x}+\beta \frac{\partial^3 u}{\partial x^3}+\gamma \frac{\partial^2 u}{\partial x^2}+\delta u=0
\end{equation}
for which the case $n=1$ represents  the KdV-Burgers equation \cite{KdVB}, while  $n=2$ corresponds to the Gardner equation (sometimes known as the modified KdV-Burgers),  which is  the equation for nonlinear dissipative waves in solid wave guides \cite{Gard}.  The terms that contain the  $\gamma$ and $\delta$ coefficients take an account the effects of  dissipation and  an external linear perturbation, and are imperative for leading to chaos.  Such additional terms appear in  non-uniform plasmas in the form of viscous or collisional effects, and they have  been introduced as a perturbation of the KdV equation by Allen and Rowlands  \cite{Allen}, where they have found the form and speed of a new solitary-wave solution to the  perturbed KdV equation that depends on two parameters. The case $\gamma=0$ has been discussed by Karpman and Maslov where they  found that the original soliton changes speed and shape, and forms a long tail at the end of which there are small oscillations in time and space\cite{Karp}. In  Chang et al. the authors  analyzed the behavior of an ion-acoustic soliton as it propagates from a high density to a low density region  \cite{Chang}, while   Kivshar and Malomed \cite{Kiv}  utilized the  inverse scattering technique to establish  the evolution of the single soliton solution in the presence of perturbations. Exact solutions to \eqref{eq1} when $\gamma=0$ and $\delta=0$ have been obtained  using the factorization method by Cornejo-P\'erez and collaborators \cite{Har}. 

In this paper we discuss the case where both coefficients $\gamma \ne 0$ and $\delta \ne 0$, and note that  $\alpha, \beta$ and $\gamma$  may be set to unity by using  the scaling
$u(x,t) = A U(X,T)$ with $X = B x$, and  $T = C t$.
For $A^n = \frac{\gamma^2}{\alpha \beta}$, $B = \frac{\gamma}{\beta}$,
$C = \frac{\gamma^3}{\beta^2}$,
one obtains the non-dimensional equation
\begin{equation}
\label{eq2}
\frac{\partial U}{\partial T} + U^n \frac{\partial U}{\partial X} + \frac{\partial^2 U}{\partial X^2}+ \frac{\partial^3 U}{\partial X^3}+\epsilon U = 0,
\end{equation}
with  $\epsilon = \frac{\delta}{C}=\frac{\delta \beta^2}{\gamma^3} \ne0.$ 
Employing the traveling wave ansatz $U(\xi) = U(X-\text{v} T)$, where $\xi$ is the traveling wave variable, $\text{v}$ is translational speed  of waves traveling in the positive $X$ direction at time $T$, and is a bifurcation parameter, then (\ref{eq2}) becomes
\begin{equation}
\label{eq3}
\frac{d^3 U}{d \xi^3}+\frac{d^2 U}{d \xi^2}+(U^n-\text{v})\frac{d U}{d \xi}+\epsilon U=0.
\end{equation}
As a  three-mode dynamical system  \eqref{eq3} takes the compact form
\begin{equation}\label{system1}
\frac{dz}{d\xi}=Jz+h,
\end{equation}
where  $U=U_1$, $z=(U_1,U_2,U_3)^T \in \mathbb{R}^3$, $J=\begin{pmatrix} 0 & 1 & 0\\ 0 & 0 & 1\\ -\epsilon & \text{v} & -1 \\\end{pmatrix}$ is the Jacobian matrix, and $ h=(0,0,-{U_1}^nU_2)^T$.

\section{Linear stability analysis}
Following standard methods of phase-plane analysis the equilibrium point $P_0\big(U_1(\xi_0),U_2(\xi_0),U_3(\xi_0)\big)$  of \eqref{system1} is the origin,  and the characteristic polynomial of the Jacobian matrix at $P_0$  is given by the cubic polynomial
\begin{equation}\label{char_eq1}
\Lambda(\lambda)=\lambda^3+\delta_1 \lambda^2+\delta_2\lambda+\delta_3,
\end{equation}
where $\delta_1=1, \delta_2=-\text{v}$, and $\delta_3=\epsilon$. 
By the Routh-Hurwitz criteria \cite{Hurwitz, Routh} the necessary and sufficient condition for the stability of the equilibrium point  $P_0$  is that all eigenvalues of (\ref{char_eq1}) must have negative real parts, which implies
\begin{equation}\label{4.7}
(\mathcal{C}_1): ~\delta_1>0,\quad\delta_3>0,\quad\delta_1\delta_2-\delta_3>0.
\end{equation}
On the contrary, one may have the onset of instability of the  solution occurring in one of two ways \cite{Mancas}. First, the equilibrium becomes non-hyperbolic due to one of the eigenvalues crossing through the origin and would occur when 
\begin{equation}\label{4.8}
 (\mathcal{C}_2):~\delta_3=0, 
 \end{equation}
which is known as  the static instability. However, this condition cannot be achieved since $\delta_3 =\epsilon \ne0$. Second, the dynamic instability condition  occurs when a pair of eigenvalues of the Jacobian become purely imaginary. The consequent Hopf bifurcation curve is given by 
\begin{equation}\label{4.9}
 (\mathcal{C}_H): \quad \delta_1\delta_2-\delta_3=0,
\end{equation}
and leads to periodic solutions of (\ref{system1}) which dissipate by  slowly converging to the origin. These solutions, denoted by $U_1(\xi), U_2(\xi)$ and $U_3(\xi)$ may be asymptotic to the origin or to a limit cycle depending on the super- or subcritical nature of the bifurcation, and in general are quasi-periodic wavetrains, since their periods are  typically incommensurate \cite{Mancas}. 

To find the eigenvalues  of the Jacobian matrix, we shift the eigenvalue according to  $\lambda=\mu -\frac{1}{3}$ to obtain  the depressed cubic 
\begin{align}\label{depressed}
\mu^3+p\mu+q=0,
\end{align}
with coefficients $p=-\text{v}-\frac 1 3$ and $q=\epsilon +\frac {\text{v}}{ 3} +\frac{2}{27}$. By  using Cardano's formula \cite{Car}, we  define 
\begin{equation}\label{system1bis}
\begin{array}{l}
X=\sqrt[3]{-\frac{q}{2}+\sqrt{\Delta}},\\
Y=\sqrt[3]{-\frac{q}{2}-\sqrt{\Delta}},
\end{array}
\end{equation}
where the modular discriminant is $\Delta=\frac{q^2}{4}+\frac{p^3}{27}$.
For a real non-repeated eigenvalue and a complex conjugate pair, we require the additional condition  
\begin{equation}\label{4.10}
(\mathcal{C}_\Delta): ~\Delta>0, 
\end{equation}
which gives the eigenvalues 
\begin{equation}\label{system2}
\begin{array}{ll}
\lambda_1&=X+Y-\frac{1}{3},\\
\lambda_{2,3}&=-\frac{1}{2}\left(X+Y\right)-\frac{1}{3}\pm i\frac{\sqrt{3}}{2}\left(X-Y\right).
\end{array}
\end{equation}
This condition  leads to the elliptic curve inequality 
\begin{equation}
g(\text{v},\epsilon)=4{\text{v}}^3+{\text{v}}^2-18\epsilon {\text{v}}-\epsilon(4+27 \epsilon)<0,
\end{equation}
 where we set  $g(\text{v},\epsilon)=-108 \Delta$.

\begin{figure}[ht!]
\centering 
\includegraphics[width=0.6\textwidth]{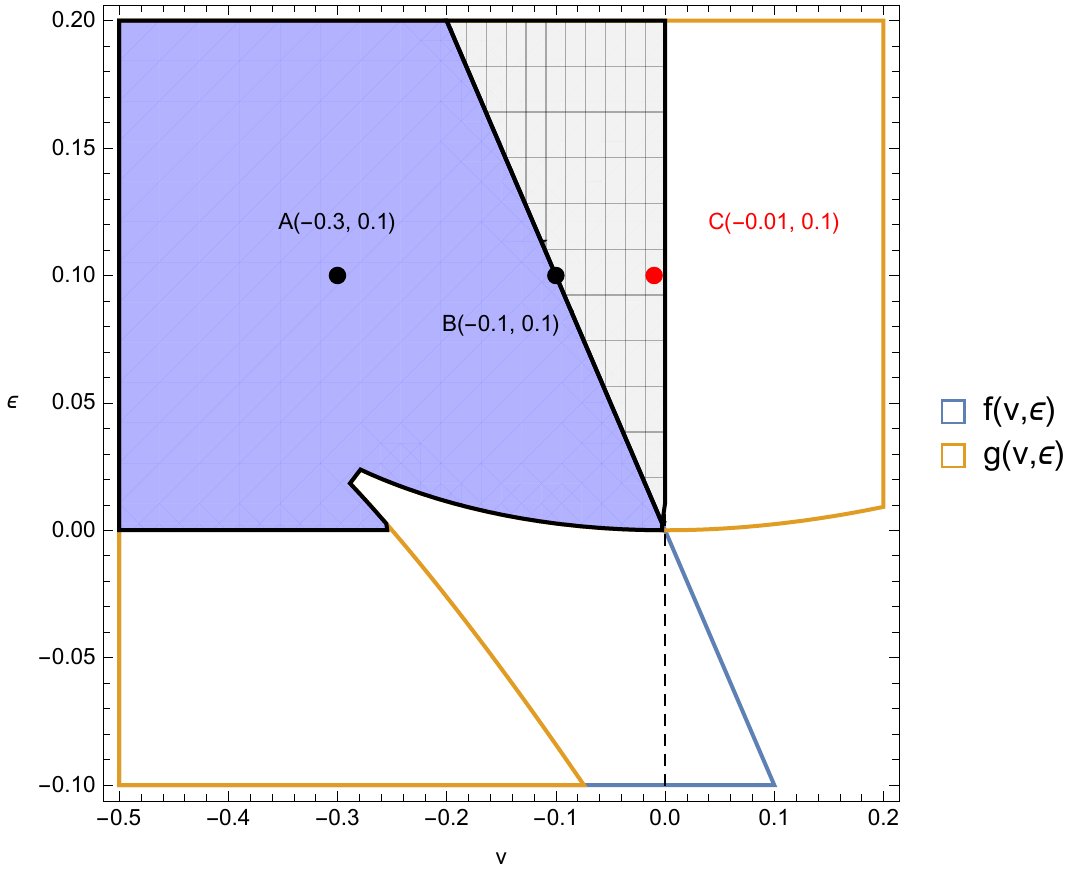}
\caption{The shaded region where point A is located is the region where  stable solutions of (\ref{system1}) exist. The point B is located on the line where  the Hopf bifurcation occurs  which  separates it from the  region where the point C is located and  unstable solutions exist. }
\label{figure1}
\end{figure}

We summarize all these conditions as follows: 
\begin{enumerate}
\item [i)] the  Hopf bifurcation curve $(\mathcal{C}_H)$ is given by  $$ f(\text{v},\epsilon)= \text{v}+\epsilon=0$$ and leads to periodic solutions,
\item [ii)]  for stability we require that   $(\mathcal{C}_1)$  holds which gives 
$$f(\text{v},\epsilon)<0\iff 0<\epsilon<-\text{v},$$ 
\item[iii)] using  $(\mathcal{C}_\Delta)$ we also require  that $g(\text{v},\epsilon)<0$.
\end{enumerate}
All these three conditions  combined depict the  dark shaded region of  Fig. \ref{figure1}.
 
For a supercritical Hopf bifurcation the complex eigenvalues cross the imaginary axis into the right half plane, and the Lyapunov coefficient calculated using \eqref{Lyap_exp} is negative. On the boundary of stability the eigenvalues are $\lambda_1<0$, and $\lambda_{2,3}=\pm i\omega_0$ for $\omega_0>0$. These eigenvalues determine  a factorization of the characteristic polynomial (\ref{char_eq1}) as 
\begin{align}\label{char_eq2}
(\lambda-\lambda_1)(\lambda^2+{\omega_0}^2)=0,
\end{align}
which leads to $\lambda_1=-1$, $\text{v}=-{\omega_0}^2$, and $\epsilon={\omega_0}^2$. The last two relations give the condition $(\mathcal{C}_H)$, and using (\ref{system2}), on the Hopf curve we have 
\begin{equation}\label{system3}
\begin{array}{l}
X+Y=-\frac 2 3,\\
X-Y=\frac {2}{\sqrt 3}\omega_0,
\end{array}
\end{equation}
which gives the solutions $X=-\frac 1 3 (1-\sqrt 3 \omega_0)$, and  $Y=-\frac 1 3 (1+\sqrt 3 \omega_0)$. 

Now we will perturb  the manifold $(\mathcal{C}_H)$ by a small velocity $\text{v}_0$, such that $\epsilon=-\text{v}+\text{v}_0$. Thus, if $\text{v}_0>0$ we are located on the right side of  $(\mathcal{C}_H)$,  while if $\text{v}_0<0$, we are on the left. Returning to the complex eigenvalues we have $\lambda_{2,3}(\text{v},\epsilon)=\nu( \text{v},\epsilon)\pm i\omega(\text{v},\epsilon)$, with $\nu(\text{v},-\text{v})=0$ and $\omega(\text{v},-\text{v})=\omega_0$ on $(\mathcal{C}_{H})$ which gives $\lambda_{2,3}(\text{v},\text{v}_0)=\nu(\text{v},\text{v}_0-\text{v})\pm i\omega(\text{v},\text{v}_0-\text{v})$. Along this codimension-one manifold 
we consider the following non-degeneracy conditions:
\begin{enumerate}
\item[] {(A1):} ~$\mathcal{C}_4=\left.\frac{d}{d\text{v}_0}\text{Re}~\lambda_{2,3}(\text{v},\text{v}_0)\right|_{\text{v}_0=0}\neq0,$
\item[]{(A2):} ~$\mathcal{C}_5<0$, which is the first Lyapunov coefficient given by \eqref{Lyap_exp}.
\end{enumerate}

To verify (A1) we compute $\mathcal{C}_4$, and we have
\begin{equation}
\left. \frac{d}{d\text{v}_0}\nu(\text{v},\text{v}_0-\text{v})\right|_{\text{v}_0=0}=
\frac{\sqrt[3]{-1+9\text{v}+27 \sqrt{\tilde \Delta}}-\sqrt[3]{-1+9\text{v}-27 \sqrt{\tilde \Delta}}}{36 \sqrt{\tilde \Delta}}\neq 0
\end{equation}
where $\tilde{\Delta}=\Delta \left.\right|_{\epsilon=-\text v}=\frac{1}{27}(-\text v)(\text v-1)^2>0$.

\begin{figure}[ht!]
\centering
\includegraphics[width=0.6\textwidth]{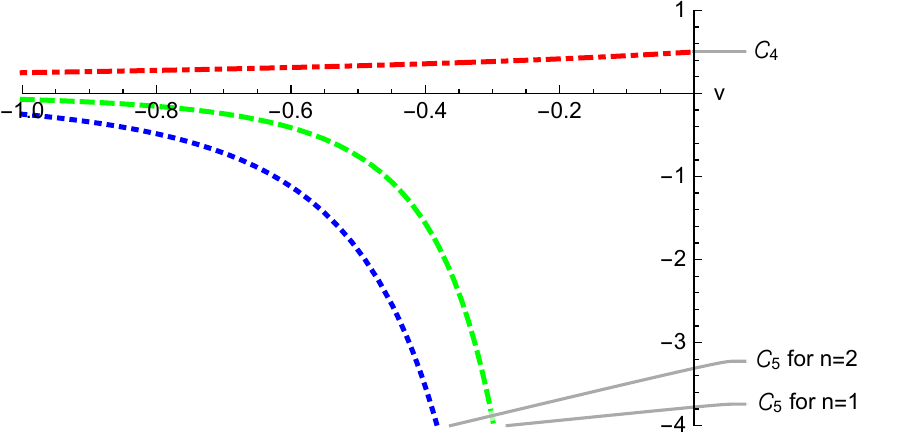}
\caption{The graphs of $\mathcal{C}_4$ and  Lyapunov coefficient $\mathcal{C}_5$  of the non-degeneracy conditions   (A1) and (A2)  which hold along ($\mathcal C_H$) and show that system \eqref{system1} undergoes a supercitical bifurcation for both KdV-Burgers and Gardner equations.}
\label{figure2}
\end{figure}

To verify (A2) we now need to  determine $\mathcal{C}_5$  which is dependent on the Taylor expansion of $h(z)$ about $z=0$. This  can be written up to order four as $h(z)=\frac{1}{2}B(z,z)+\frac{1}{6}C(z,z,z)+O(\left\|z\right\|^4)$, where $B(a,b)$ and $C(a,b,c)$ are multilinear vector functions with components $B_j, C_j$ defined by the vectors $a,b,c \in \mathbb{R}^3 $  and are given by
\begin{equation}
\begin{array}{ll}
B_j(a,b)&=\sum^{3}_{k,l=1}\left.\frac{\partial^2 h_j}{\partial U_{k}\partial U_{l}}\right|_{U=0}\;a_kb_l, \\
C_j(a,b,c)&=\sum^{3}_{k,l,m=1}\left.\frac{\partial^3 h_j}{\partial U_{k}\partial U_{l}\partial U_{m}}\right|_{U=0}\;a_kb_lc_m,
\end{array}
\end{equation} 
 for $j=1,2,3$.  
Along the curve $(\mathcal{C}_{H})$, $J$ has a pair of purely imaginary eigenvalues  $\lambda_{2,3}=\ \pm i \omega_0$ with corresponding eigenvectors  $r_{2,3}$ such that $Jr_{2,3}=\pm i\omega_0 r_{2,3}$.  Writing these values in terms of $\text{v}$ as $\lambda_{2,3}=\pm i \sqrt{- \text{v}}$, we have the eigenvectors 
$$
r_{2,3}=\left(\frac{1}{\text {v}},\frac{\mp i}{\sqrt{-\text{v}}},1\right)^T.
$$
Let $s_{2}$ be the corresponding adjoint eigenvector such that  $J^{T}s_{2}=- i\omega_0 s_{2}$, and since we require that $\left\langle s_2,r_2\right\rangle=1$, then 
$$s_{2}=\left(\frac{-\text{v}(1-i \sqrt{-\text{v}})}{2(1-\text{v})},\frac{-2\text{v}+i\sqrt{-\text{v}}(\text{v}+1)}{2(1-\text{v})},\frac{-\text{v}+ i \sqrt{-\text{v}}}{2(1-\text{v})}\right)^T.$$
Now, we can then determine the expression for $\mathcal{C}_5$ as given by Kuznetsov \cite{Kuz},  which is
\begin{equation}
\begin{array}{ll}\label{Lyap_exp}
\mathcal{C}_5&=\frac{1}{2\sqrt{-\text{v}}}\text{Re}\left[\left\langle s_2,C(r_2,r_2,r_3)\right
\rangle-2\left\langle s_2,J^{-1}B(r_2,J^{-1}B(r_2,r_3))\right \rangle \right. \\
&+\left .\left\langle s_2,B(r_3,(2i\sqrt{-\text{v}} I-J)^{-1}B(r_2,r_2))\right\rangle\right],\
\end{array} 
\end{equation}
where $I$ is the identity matrix. 
This expression depends on $B(a,b)$ and $C(a,b,c)$ which also depend on $n$. 
\begin{enumerate}
\item{For  $n=1$} we have 
$$B(a,b)=\left(0,0,-(a_1b_2+a_2b_1)\right)^{T}$$
while $C(a,b,c)$ is zero, and by using  (\ref{Lyap_exp}) the Lyapunov coefficient is
\begin{equation}\label{lyap}
\begin{array}{ll}
\mathcal{C}_5&=\frac{1}{2\sqrt{-\text{v}}}\text{Re}\left\langle s_2,B(r_3,(2i\sqrt{-\text{v}} I-J)^{-1}B(r_2,r_2))\right\rangle\\
&=-\frac{1}{2\text{v}^2\sqrt{-\text{v}}(1-5\text{v}+\text{v}^2)}.
\end{array}
\end{equation}
\item{For $n=2$} we have 
$$C(a,b,c)=(0,0,-2(a_1b_2c_1+a_1b_1c_2+a_2b_1c_1))^{T}$$
while $B(a,b)$ is zero,
and  using  (\ref{Lyap_exp}) gives 
\begin{equation}\label{n2} 
\begin{array}{ll}
\mathcal{C}_5&=\frac{1}{2\sqrt{-\text{v}}}\text{Re}\left[\left\langle s_2,C(r_2,r_2,r_3)\right\rangle-2\left\langle s_2,J^{-1}B(r_2,J^{-1}B(r_2,r_3))\right\rangle\right.\\
&\left.+\left\langle s_2,B(r_3,(2i\sqrt{-\text{v}} I-J)^{-1}B(r_2,r_2))\right\rangle\right]\\
&=\frac{1}{2\sqrt{-\text{v}}}\text{Re}\left\langle s_2,C(r_2,r_2,r_3)\right\rangle\\
&=-\frac{1}{2{\text{v}}^2\sqrt{-\text{v}}(1-\text{v})}.
\end{array}
\end{equation}
\end{enumerate}
The non-degeneracy conditions (A1) and (A2) hold along the curve $(\mathcal{C}_H)$ since $\mathcal{C}_4>0$  and $\mathcal{C}_5<0$ for $\text{v}<0$, see Fig. \ref{figure2}. Thus, in both cases $n=1$ and $n=2$, \eqref{system1} undergoes a supercritical Hopf bifurcation, and hence a unique stable limit cycle bifurcates from the origin.

\begin{figure}[ht!]
  \centering
    \includegraphics[width=0.75\textwidth]{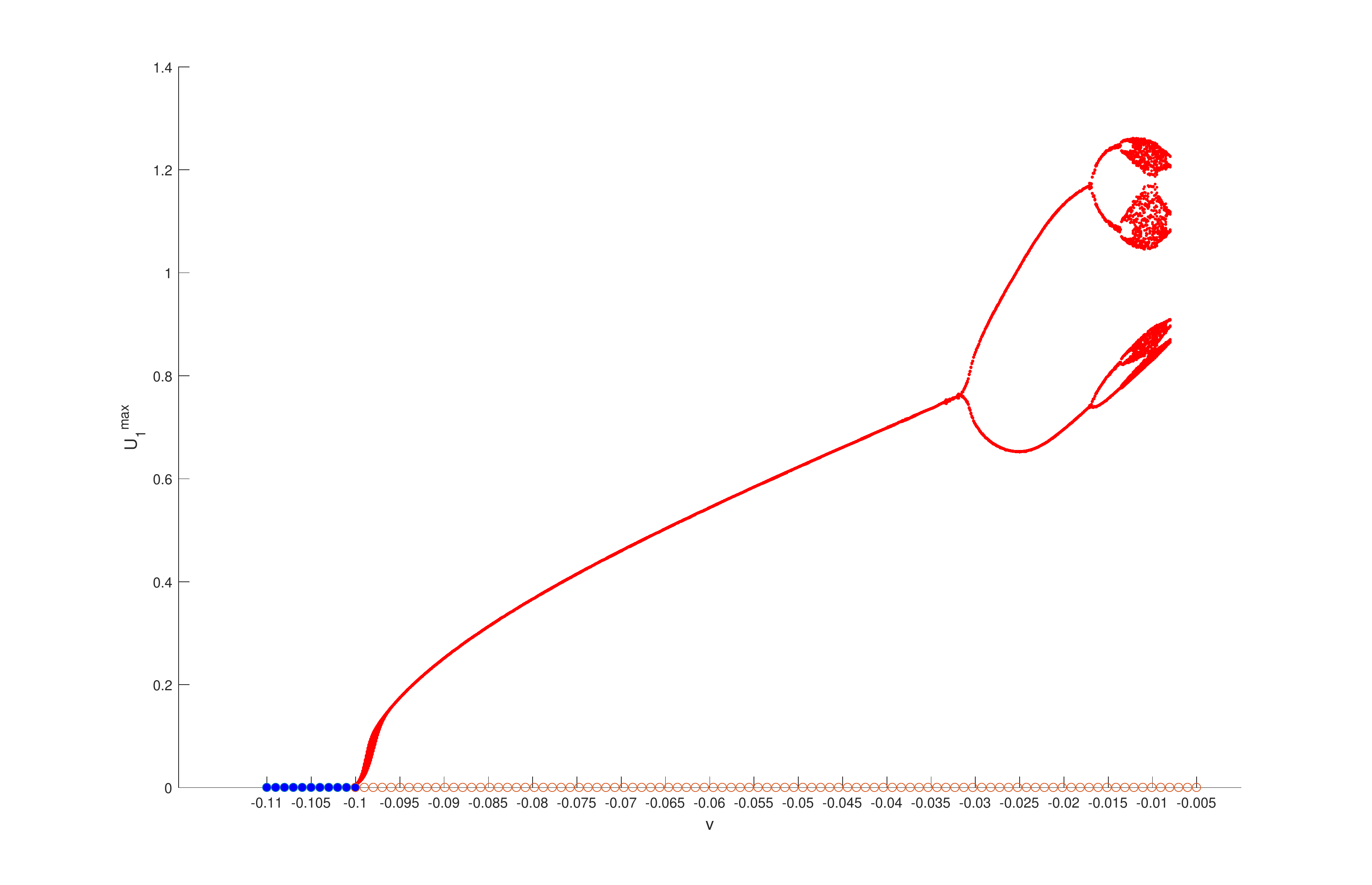}
    \caption{Bifurcation diagram for KdV-Burgers equation ($n=1$) for  $\epsilon=0.1$ and  $\text v \in [-0.11,-0.00791]$. The equilibrium point is first  stable (solid circles) then  unstable (open circles). At $\text v=-0.1$ we have a change of stability while around $\text v=-0.03$ we witness  period doubling. As $\text v$  varies even more to the right  the cascade of bifurcations  will eventually lead to chaos.}
		\label{figure3a}
		\end{figure}

\begin{figure}[ht!]
  \centering
        \includegraphics[width=0.75\textwidth]{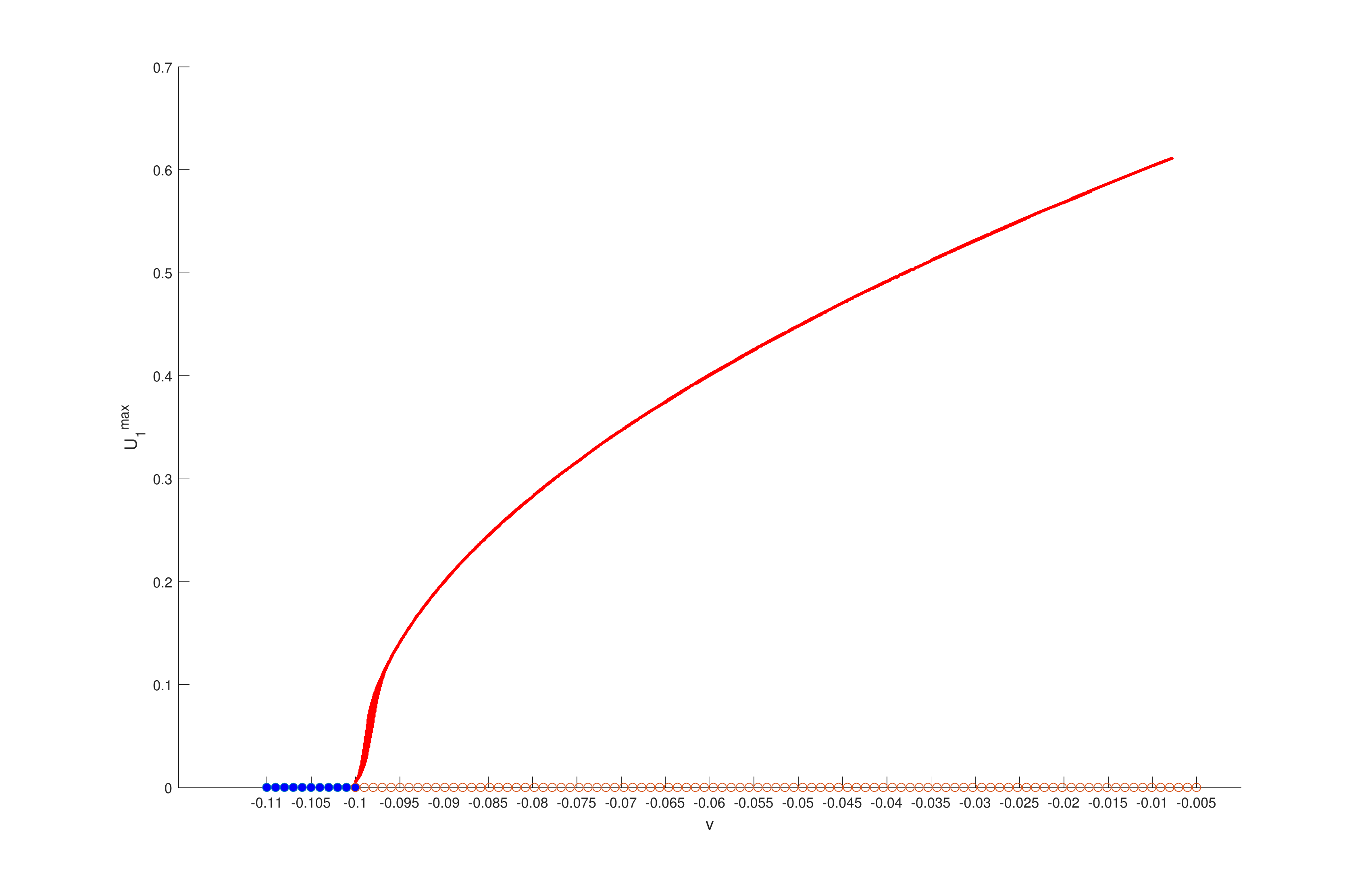}
    \caption{Bifurcation diagram for Gardner  equation ($n=2$) for $\epsilon=0.1$ and $\text v \in [-0.11,-0.00791]$. Here we have only a change of stability and no chaos is present.}
		\label{figure3b}
		\end{figure}

Now that we have established that \eqref{system1} undergoes a supercritical Hopf bifurcation, we turn our attention to specific points in the parameter space of Fig. \ref{figure1} which will uncover further qualitative properties of the solutions to \eqref{system1}. For a fixed $\epsilon=0.1$, on the left of the Hopf curve  ($\mathcal{C}_H)$  for $\text{v}<-0.1$ the equilibrium point shown as solid circles in the bifurcation diagrams of Figs. \ref{figure3a} and \ref{figure3b}  is asymptotically stable, while the open circles in the same figures indicate that the equilibrium point is unstable.  For case A when $\text v=-0.3$, the dissipative periodic orbit is shown in  Fig. \ref{figA09} of the Appendix. Because case A is located to the left of the Hopf curve, the solution converges asymptotically to the origin.  On the Hopf curve $\text{v}=-0.1$, the origin becomes weakly stable, this is witnessed by considering case B and the corresponding solution in Fig. \ref{figB09} of the Appendix. Passing to the right of the Hopf curve for $\text{v}>-0.1$, the origin becomes unstable and a stable limit cycle exists for case C. When $n=1$ this limit cycle undergoes a period doubling bifurcation, see Fig. \ref{figure3a}, and the formation of limit cycles of period two, four, eight, etc. are created which is a classic signature of chaos.  The limit cycle in this case is a chaotical attractor which will be confirmed using Shil'nikov's analysis in the following section.  Whereas for $n=2$ a unique stable limit cycle is produced see Fig. \ref{figure3b}, and no chaos is observed.  We  compute numerically the solutions for case C, for both  $n=1$ and $n=2$ resepctively. The solution which tends towards a chaotical attractor for $n=1$ is shown in  Fig. \ref{figC09}, while for $n=2$, the solution which tends to a stable  periodic orbit is shown in Fig. \ref{figD09} of the Appendix.
		
Next, we compute numerically the maximal Lyapunov exponent for  $\text v \in [-0.1, -0.00791]$ for both $n=1$ and $n=2$ by using the algorithm detailed in \cite{wolf}, which measures the degree of sensitivity to initial conditions.  

\begin{figure}[ht!]
  \centering
       \includegraphics[width=0.7\textwidth]{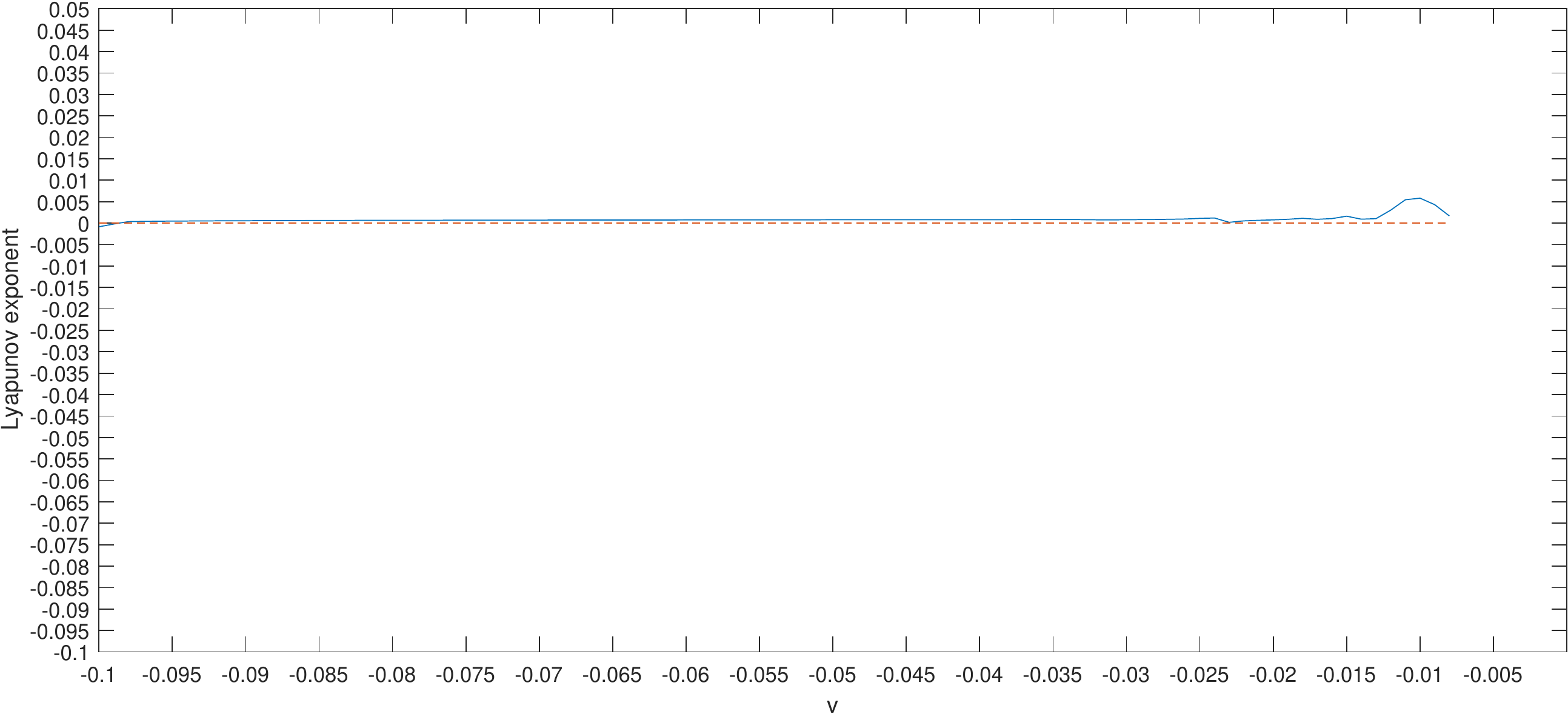}
        \caption{The graph of the  maximal Lyapunov exponent for KdV-Burgers equation ($n=1$)  for  $\epsilon=0.1$  and $\text v \in [-0.1,-0.00791]$. Notice that the exponent takes the largest positive value in the ``hump'' region around $\text{v}=-0.01$, which indicates a strong deterministic chaos in that region.}  
		\label{figurelyapa}
		\end{figure}

When $n=1$, we can infer from Fig. \ref{figurelyapa}   that as the Hopf curve is transversed in the $\text{v}$ direction at $\text v =-0.1$ the equilibrium becomes a saddle-focus and eventually the system experiences chaotic dynamics which is due to a period doubling bifurcation.  The graph of the largest Lyapunov exponent is  positive for $\text{v}$ around $-0.01$  and gives us a parameter range for selecting candidate values of $\text{v}$ when applying the Shil'nikov criteria. In this region the phase portraits  near the limit cycle have  an infinite number of saddle cycles  which correspond to periodic wavetrains of \eqref{system1}. The limit cycle  has  an oscillating tail, while secondary limit cycles  near the saddle-focus homoclinic bifurcation correspond to double traveling impulses of \eqref{system1}.    This  is synonymous with deterministic chaos provided the corresponding unstable manifold folds back onto itself and remains confined within a bounded domain which  confirms the existence of periodic and chaotic trajectories for KdV-Burgers equation.

\begin{figure}[ht!]
  \centering
         \includegraphics[width=0.7\textwidth]{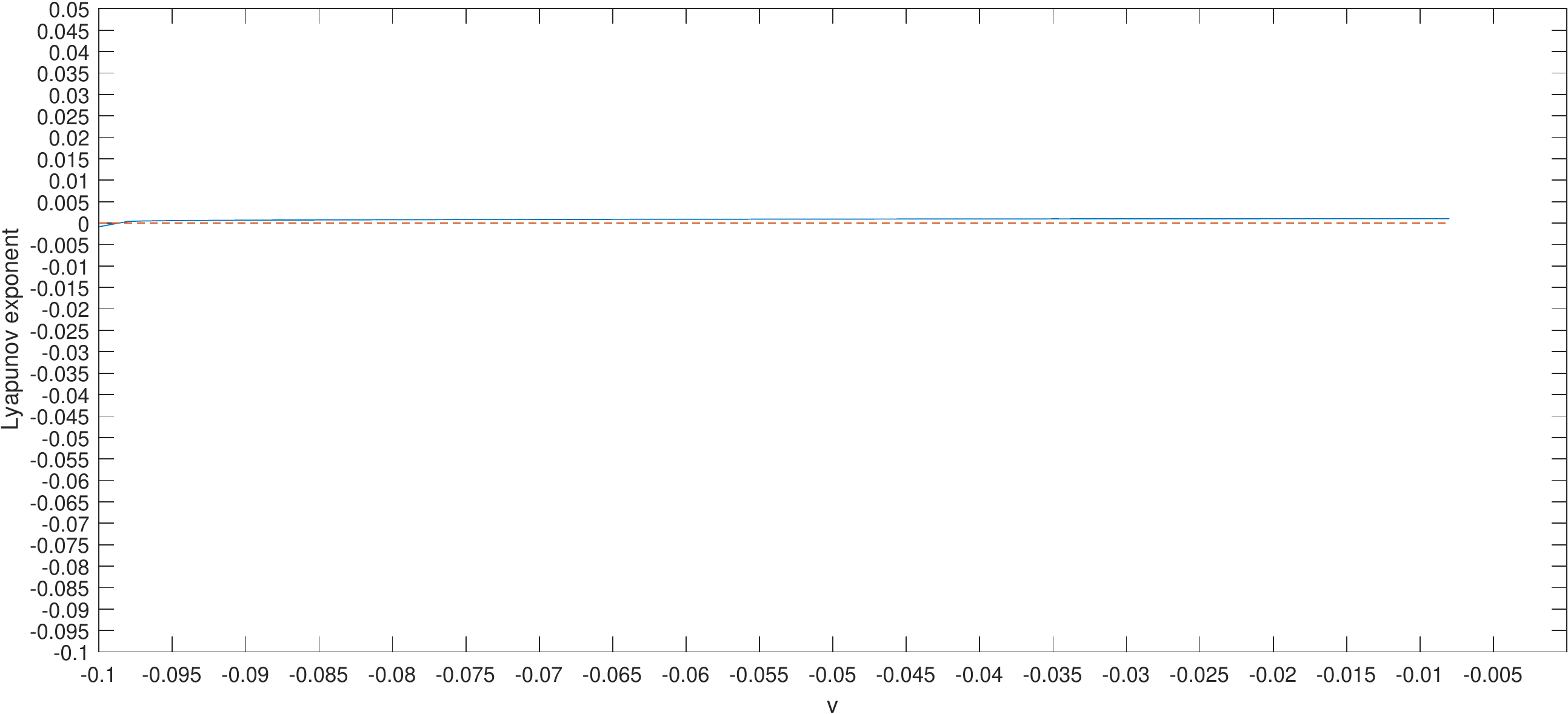}
        \caption{The graph of the  maximal Lyapunov exponent for Gardner equation ($n=2$)  for $\epsilon=0.1$ and  $\text v \in [-0.1,-0.00791]$. Notice that the exponent is very close to zero throughout the whole interval.}  
		\label{figurelyapb}
		\end{figure}

When $n=2$, Fig. \ref{figurelyapb} indicates a close to zero Lyapunov exponent, that is, no sensitivity to initial conditions. This combined with the bifurcation diagram of Fig. \ref{figure3b} leads us to conclude that there is no chaos present for \eqref{system1} in the case of Gardner equation.
  
\section{Shil'nikov's analysis of homoclinic orbits}
\begin{figure}[ht!]
\centering
\includegraphics[width=.6\textwidth]{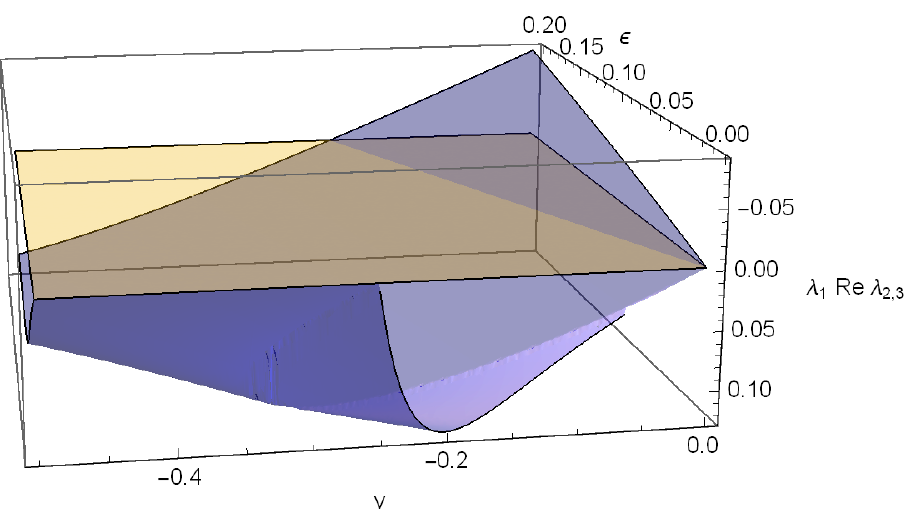}
\caption{The graph of $\lambda_1\cdot \text{Re}~\lambda_{2,3}$  for $\epsilon=0.1$ and $-0.5\leq\text{v}\leq0$.}
\label{figure7a}
\end{figure}

\begin{figure}[ht!]
\centering
\includegraphics[width=.6\textwidth]{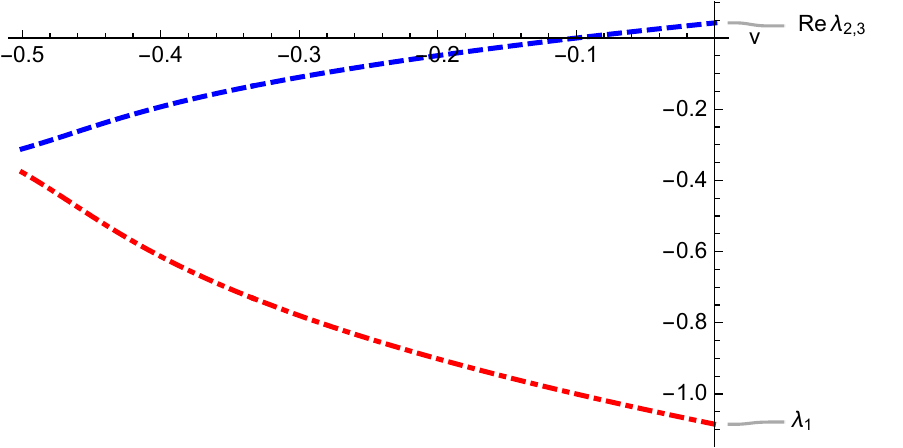}
\caption{The graph of $\lambda_{1}$ and $\text{Re}~\lambda_{2,3}$ for $\epsilon=0.1$ and $-0.5\leq\text{v}\leq0$.}
\label{figure8a}
\end{figure}

In this section, the algebraic expression of the homoclinic orbit is derived for $n=1$ in \eqref{system1}, and
the uniform convergence of its series expansion is
proved. In particular, the Shil'nikov criterion \cite{Roy} is verified  by applying Shil'nikov's theorem (Theorem~2.1 in \cite{Sil}), and thus ensuring that the system \eqref{system1} is chaotic. We fix $\epsilon=0.1$ starting from the initial condition  $(0.001,0.001,0.001)$, and we vary the bifurcation parameter $\text v$.  As observed in Fig. \ref{figure3a}, a period doubling route to chaos is present for $n=1$.  In order to carry out the Shil'nikov analysis, the conditions (B1), (B2) together with the existence of a homoclinic orbit which defines a traveling impulse   must be established. These conditions are:
 \begin{enumerate}
\item[] {(B1):}  $\lambda_1\cdot\text{Re}~\lambda_{2,3}<0$,  
\item[] {(B2):} $\left|\lambda_1\right|>\left|\text{Re}~\lambda_{2,3}\right|$ (\textit{the Shil'nikov inequality}).
\end{enumerate}
Let us first consider (B1),  for  $-0.5\leq\text{v}\leq0$, and $-\text{v}<\epsilon\leq0.2$, which is a region to the right of ($\mathcal C_H$) depicted by the meshed area in Fig. \ref{figure1} where point C is located. Since $\Delta>0$ we can explicitly compute $\lambda_1\cdot\text{Re}~\lambda_{2,3}$. From Fig. \ref{figure7a}, we see that to the right of the Hopf curve the origin becomes a saddle focus, therefore (B1) is satisfied.  For (B2) we have the following relation between eigenvalues: $\text{Re}~\lambda_{2,3}=-\frac{1}{2}\lambda_1-\frac{1}{2}$,  therefore (B2) can be expressed as $-2\lambda_1>\left|\lambda_1+1\right|$.  This inequality is easily verified when $\lambda_1+1<0$.  On the other hand if $\lambda_1+1>0$, then we have  $\lambda_1<-\frac{1}{3}$. As we can see  from Fig. \ref{figure8a} this is  valid for $\epsilon=0.1$ and $-0.5\leq\text{v}\leq0$.
		
We now discuss the existence of a non-trivial homoclinic orbit of \eqref{system1} connecting the origin to itself, that is, a trajectory $z=(U_1,U_2,U_3)^T$ of \eqref{system1} (not restricted to the origin) which is doubly asymptotic to the origin.  We use the parameter values given by case C to construct the homoclinic orbit, but  the procedure carried out below can be adapted for other parameter regimes as well.
\begin{theorem}
If $\Delta>0$ and (B1), (B2) are satisfied, then system \eqref{system1} has one homoclinic orbit whose first component has the form \eqref{hom_sol}, and the corresponding chaos is of horseshoe type.
\end{theorem}
\proof
We start by assuming a solution to \eqref{eq3} is in the form 
\begin{align}\label{hom_sol}
U_1(\xi)=\sum^{\infty}_{k=1}b_{k}e^{k\rho \xi}
\end{align}
for $\xi>0$, where $\rho$ is an undetermined constant with $\text{Re}~\rho<0$.   From \eqref{eq3} with $n=1$ we have the following relation
\begin{align}\label{exp_rel}
\sum^{\infty}_{k=1}\left[k^3\rho^3+k^2\rho^2-\text{v}k\rho+\epsilon\right]b_{k}e^{k\rho\xi}+\rho\sum^{\infty}_{k=2}\left[\sum^{k-1}_{i=1} b_{k-i}b_{i}(k-i)\right]e^{k\rho\xi}=0.
\end{align}
Comparing the coefficients of $e^{k \rho \xi}$ in \eqref{exp_rel} of the same power terms for $k=1$ we obtain 
\begin{equation}\label{bet}
\rho^3+\rho^2-\text{v}\rho+\epsilon=0,
\end{equation}
which is equivalent to $\Lambda(\rho)=0$.  In the meshed region of  Fig. \ref{figure1}, where point C is located, and where the condition (B1) is satisfied we know that $\Delta>0$, therefore a unique negative root of $\Lambda(\rho)=0$ is determined by the real eigenvalue $\lambda_1$ in \eqref{system2}. Notice that in this region the two other roots $\lambda_{2,3}$ have positive real parts, see  Fig. \ref{figure8a}.  Hence we restrict our analysis to $\rho=\lambda_1$ since using any of the other two eigenvalues would lead to a divergent series.

To retrieve asymptotically the non-trivial homoclinic orbit we must assume that in \eqref{hom_sol} $b_1\neq 0$, otherwise when $b_1=0$ then $b_k=0$ for all $k\geq2$.  Now if $b_1\neq0$, for $k=2,3$ we have the following values 
\begin{equation}
\begin{array}{ll}
b_2=-\frac{{b_1}^2}{7\rho^2+3\rho-\text{v}},\\
b_3=\frac{3{b_1}^3}{(7\rho^2+3\rho-\text{v})(26\rho^2+8\rho-2\text{v})}.
\end{array}
\end{equation}
In general for $k\geq2$ we have the recursive formula
\begin{equation}\label{rec_rel1}
b_{k}=-\frac{\rho}{\Lambda (k\rho)}\sum^{k-1}_{i=1}b_{k-i}b_{i}(k-i).
\end{equation}
Using \eqref{bet} then  \eqref{rec_rel1} becomes 
\begin{equation}\label{rec_rel}
b_{k}=-\frac{1}{h_k(\rho)}\sum^{k-1}_{i=1}b_{k-i}b_{i}(k-i),
\end{equation}
where
$$h_{k}(\rho)=(k^3-1)\rho^2+(k^2-1)\rho-(k-1)\text{v}.$$

From \eqref{rec_rel} we see that the coefficients $b_k$ can be written in the closed form \begin{equation}\label{close}
b_k=L_k{b_1}^k,  \quad  \text{for}\;\; k \geq 2
\end{equation}
where the $L_{k}$'s can be computed by induction on $k$ and can be seen as functions depending upon $\epsilon$, $\text{v}$, and $\rho$.  

To find the expression of $U_1(\xi)$ for $\xi<0$ we assume that 
\begin{align}
U_1(\xi)=\sum^{\infty}_{k=1}a_{k}e^{-k \rho \xi}.
\end{align}
 As in the case $\xi>0$, we have 
\begin{equation}\label{close2}
a_k=M_k{a_1}^k,   \quad \text{for}\;\; k \geq 2
\end{equation}
where the $M_{k}$'s are some other functions that also depend on $\epsilon$, $\text{v}$, and $\rho$ and  satisfy $M_k=-L_{-k}$. If  $a_1=-b_1$ then  
\begin{equation}\label{ais}a_k=M_k{a_1}^k=(-1)^{k+1}L_{-k}{b_1}^k=(-1)^{k+1}b_{-k}{b_1}^{2k},  \quad  \text{for}\;\; k \geq 2.
\end{equation}
In order to construct a component of the homoclinic orbit we must impose the following smoothness conditions
\begin{align}
U_1(0_{-})=U_1(0_{+}),\quad \frac{d{U_1}}{d \xi}\Big|_{\xi=0_{-}}=\frac{d{U_1}}{d \xi}\Big|_{\xi=0_{+}},\quad \frac{d^2{U_1}}{d \xi^2}\Big|_{\xi=0_{-}}=\frac{d^2{U_1}}{d \xi^2}\Big|_{\xi=0_{+}}.
\end{align}
Hence, the  component $U_1(\xi)$ of the homoclinic orbit $z(\xi)$ takes the form
\begin{align}\label{homoclinic_sol}
U_1(\xi) = \left\{
  \begin{array}{lr}
    \sum^{\infty}_{k=1}b_{k}e^{k \rho \xi}, & \text{for}\;\; \xi>0\\
    0, & \text{for}\;\; \xi=0\\
		\sum^{\infty}_{k=1}a_{k}e^{- k \rho \xi}, & \text{for}\;\;\xi<0
  \end{array}
\right..
\end{align}
  
To determine $b_1$ numerically we use \eqref{homoclinic_sol}  which implies that  $\sum^\infty_{k=1}L_k{b_1}^k=0$. This can be solved using Newton's method by using the corresponding truncated polynomial given by
\begin{align}\label{bk_eq}
F(b_1)=L_N{b_1}^N+L_{N-1}{b_1}^{N-1}+\cdots+L_{2}{b_1}^2+b_1=0.
\end{align}
Using $N=1000$, we find $b_1\approx 10.78743$, and since $\frac{dF}{db_1}\neq 0$ for $N$ large, this excludes the possibility that $b_1$ is a multiple root. Since  $b_1$ is known, we can find $a_1$ using \eqref{ais}.

We seek to show that the series expansion in \eqref{homoclinic_sol} converges absolutely, for this we use a series expansion argument as outlined in \cite{Zhou}.  We use the parameters  of point C given by  $\epsilon=0.1$ and $\text{v}=-0.01$. For other parameter regimes the proof is similar.  Based on the values of $b_2$ and  $b_3$ along with \eqref{rec_rel} we claim that
\begin{equation}\label{clo}
\left|b_k\right|\leq\frac{1}{k}\frac{\left|b_1\right|^k}{10.8^k},
\end{equation}
for  $k \ge k_0$, for some $k_0(\epsilon,\text{v})>0$.  For the aforementioned values of $\epsilon$ and $\text{v}$ we use Cardano's formula \eqref{system2} to determine $\rho\approx-1.07694$.  Next we compute $k=k_0=216$ by utilizing \eqref{close}
in conjunction with \eqref{clo},  
which gives the  inequality 
\begin{equation}
\left|L_k\right|\leq \frac{1}{k~10.8^k}.
\end{equation}
For $k_0=216$, we have the inequality
\begin{equation}
\left|b_{k_0}\right|\approx2.7465\times10^{-k_0-7}{\left|b_1\right|}^{k_0}\leq\frac{1}{k
_0}\frac{\left|b_1\right|^{k_0}}{10.8^{k_0}}.
\end{equation}
This estimate is established by induction. Using \eqref{rec_rel} and the induction hypothesis \eqref{clo}, we have
\begin{equation}
\begin{array}{ll}
\left|b_{k+1}\right|&\leq \frac{1}{h_{k+1}(\rho)}\sum^{k}_{i=1}\left|b_{k+1-i}\right|\left|b_{i}\right|(k+1-i)\\
&\leq\frac{1}{h_{k+1}(\rho)}\sum^{k}_{i=1}\frac{\left|b_1\right|^{k+1-i}}{(k+1-i)~10.8^{k-i+1}}\frac{\left|b_1\right|^i}{i~10.8^i}(k+1-i)\\&
=\frac{\left|b_1\right|^{k+1}}{10.8^{k+1}}\frac{1}{h_{k+1}(\rho)}\sum^{k}_{i=1}\frac{1}{i}\\
&\leq\frac{\left|b_1\right|^{k+1}}{10.8^{k+1}h_{k+1}(\rho)}\left(1+\ln(k)\right)\\
&\leq\frac{\left|b_1\right|^{k+1}}{10.8^{k+1}h_{k+1}(\rho)}k. 
\end{array}
\end{equation}
Since 
\begin{equation}
\frac{k}{h_{k+1}(\rho)}\leq\frac{1}{k+1} ,\quad \text{for}\;\;  k\geq k_0,
\end{equation} then 
\begin{equation}\label{clo2}
\left|b_{k+1}\right|\leq \frac{1}{k+1}\frac{\left|b_1\right|^{k+1}}{10.8^{k+1}}.
\end{equation} 
 
Now, we can prove the convergence of the series
\begin{equation}
\begin{array}{ll}
\left|\sum^{\infty}_{k=1}b_ke^{k \rho \xi}\right|&\leq \sum^{\infty}_{k=1}\left|b_k\right|e^{k \rho \xi}\\
&=\sum^{k_0}_{k=1}\left|b_k\right|e^{k \rho \xi}+\sum^{\infty}_{k=k_0+1}\left|b_k\right|e^{k \rho\xi}\\
&\leq\sum^{k_0}_{k=1}\left|b_k\right|e^{k \rho \xi}+\sum^{\infty}_{k=k_0+1}\frac{1}{k}\frac{\left|b_1\right|^k}{10.8^k}e^{k \rho\xi}\\
&\leq\sum^{k_0}_{k=1}\left(\left|b_k\right|-\frac{1}{k}\frac{\left|b_1\right|^k}{10.8^k}\right) e^{k \rho \xi}+\sum^{\infty}_{k=1}\frac{1}{k}\frac{\left|b_1\right|^k}{10.8^k}e^{k \rho \xi}\\
&\leq \sum^{k_0}_{k=1}\left(\left|b_k\right|-\frac{1}{k}\frac{\left|b_1\right|^k}{10.8^k}\right)e^{k\rho \xi}-\ln\left(1-\frac{\left|b_1\right|e^{k \rho \xi}}{10.8}\right)\\
&\leq\sum^{k_0}_{k=1}\left|\left|b_k\right|-\frac{1}{k}\frac{|b_1|^k}{10.8^k}\right|-\ln\left(1-\frac{\left| b_1 \right|}{10.8}\right).
\end{array}
\end{equation}

The above estimate holds for all $\xi\geq0$, thus, the series expansion \eqref{hom_sol} is uniformly convergent for all $\xi\geq0$.  The above argument can be duplicated for the case $\xi<0$, therefore \eqref{homoclinic_sol} is well defined.\eproof

\section{Conclusion}
A rigorous treatment of Hopf bifurcation analysis has been presented  for the KdV-Burgers equation for $n=1$ and the Gardner equation for $n=2$, with  dissipative $\gamma$ and  perturbation $\delta$ terms.   For both equations, left moving traveling waves are stable for $\text v< -\frac{\delta \beta ^2}{\gamma^3}$ and $\delta, \gamma$ of the same sign. For  a certain region of space  the solutions undergo a supercritical Hopf bifurcation as the wave speed $\text{v}$ crosses  the Hopf curve when  consequently a stable limit cycle is born for $\text v= -\frac{\delta \beta ^2}{\gamma^3}$.  For the KdV-Burgers equation the stable limit cycle undergoes a period doubling bifurcation indicating the existence of a Smale horseshoe.  This is verified by applying the Shil'nikov criterion.  By utilizing the undetermined coefficient method, a homoclinic orbit is identified and  explicit and convergent algebraic expressions are derived.  We show a parameter regime  for  $n=1$ when the system~\eqref{system1} has a homoclinic orbit of Shil'nikov type which implies horseshoe chaos, while for  $n=2$  in the same parameter space the supercritical Hopf bifurcation leads only to stable periodic orbits. For both equations, right traveling waves for $\text v>0$ which correspond to  dissipation and perturbation terms of opposite  sign are unstable. 
\section*{Appendix}
\begin{figure}[ht!]
  \centering
    \includegraphics[width=0.475\textwidth]{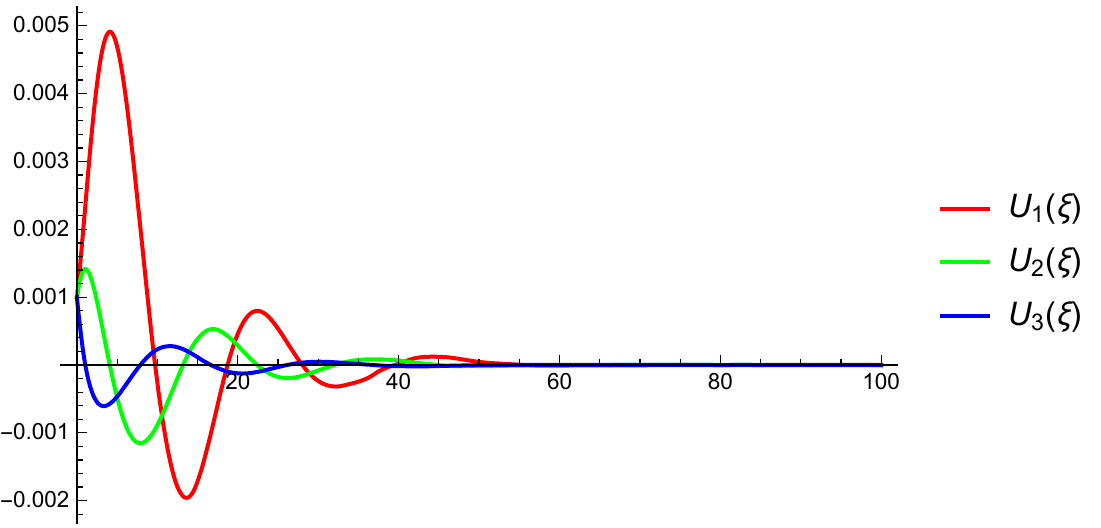}
     \includegraphics[width=0.475\textwidth]{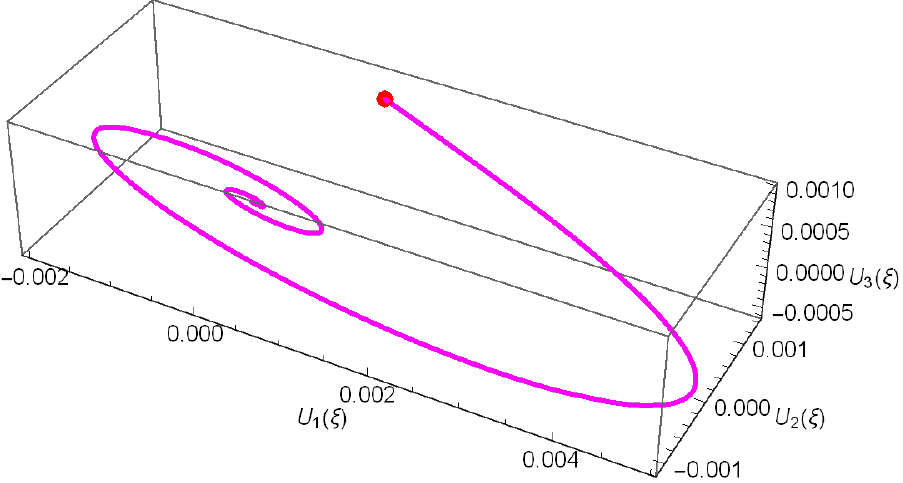}
    \caption{Case  A  for $n=1$. Stable periodic solutions (left panel). Stable periodic orbit (right panel).}
		\label{figA09}
\end{figure}

\begin{figure}[ht!]
  \centering
    \includegraphics[width=0.475\textwidth]{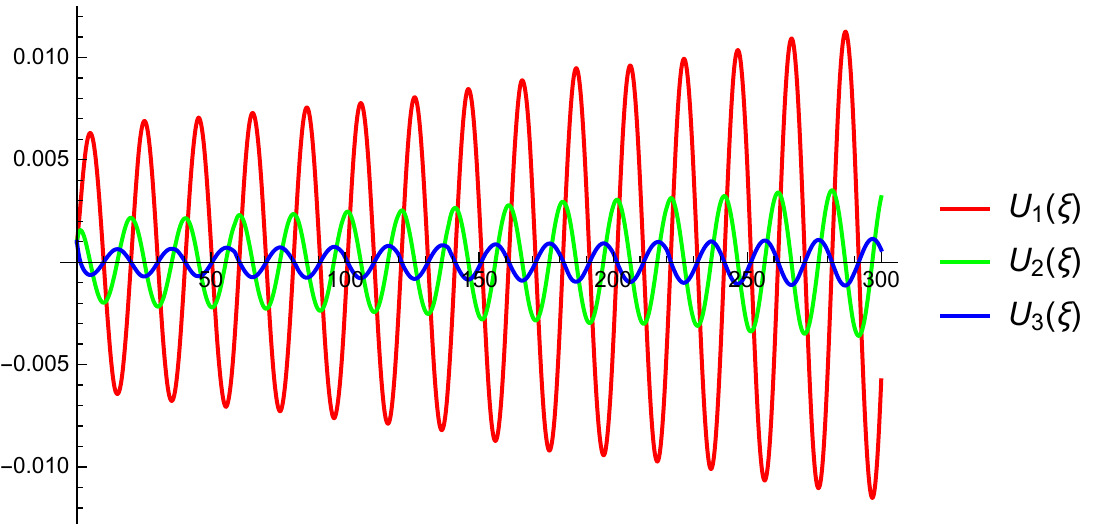}
     \includegraphics[width=0.475\textwidth]{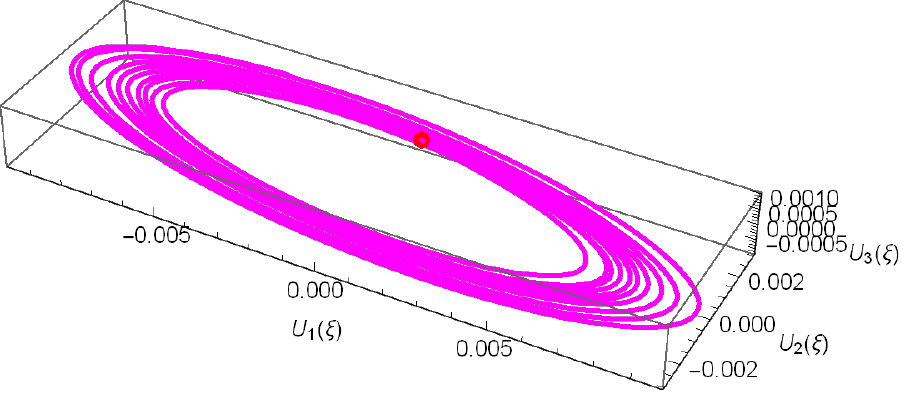}
    \caption{Case  B  for $n=1$. Periodic evolution which is weakly convergent to the origin.}
		\label{figB09}
\end{figure}

\begin{figure}[ht!]
  \centering
    \includegraphics[width=0.475\textwidth]{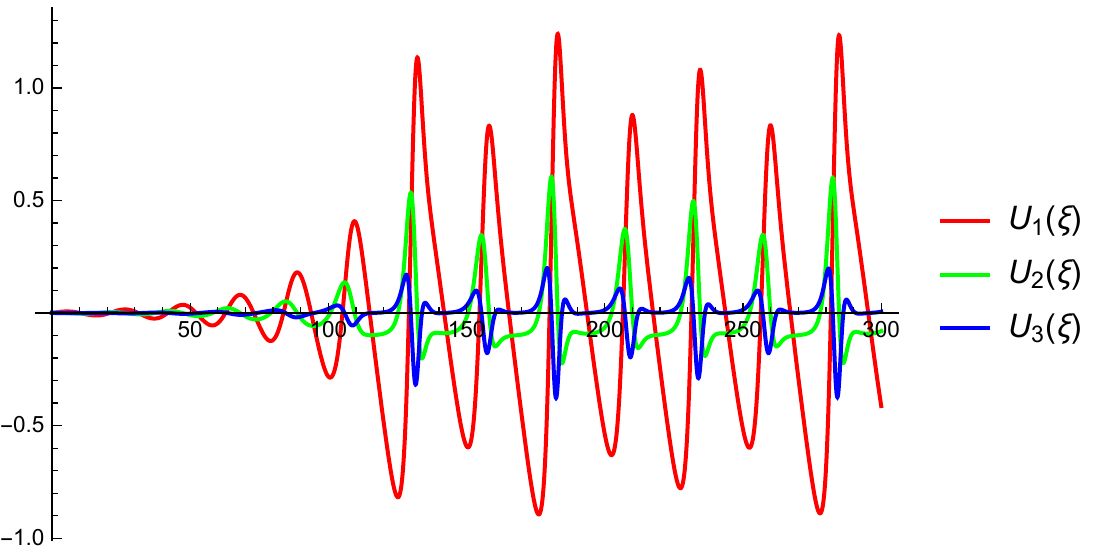}
     \includegraphics[width=0.475\textwidth]{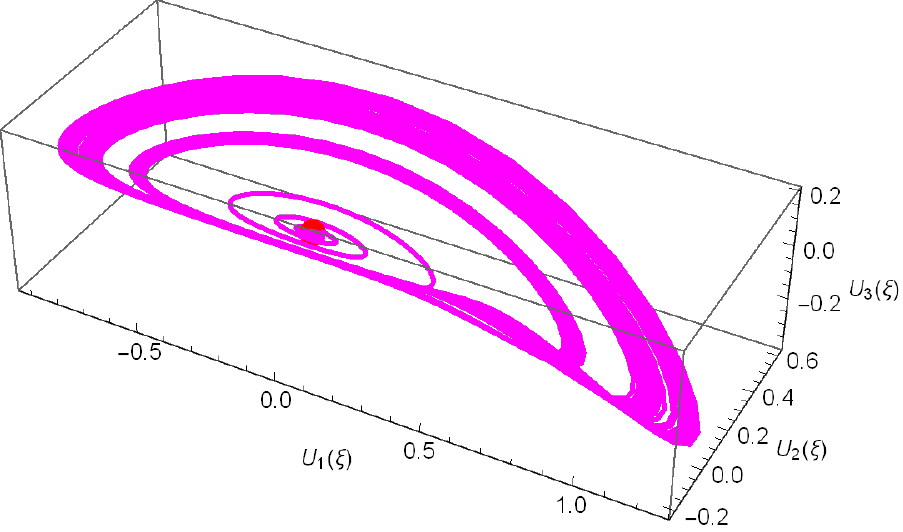}
    \caption{Case  C  for $n=1$. Chaotical evolution (left panel), chaotical attractor (right panel).}
		\label{figC09}
\end{figure}

\begin{figure}[ht!]
  \centering
    \includegraphics[width=0.475\textwidth]{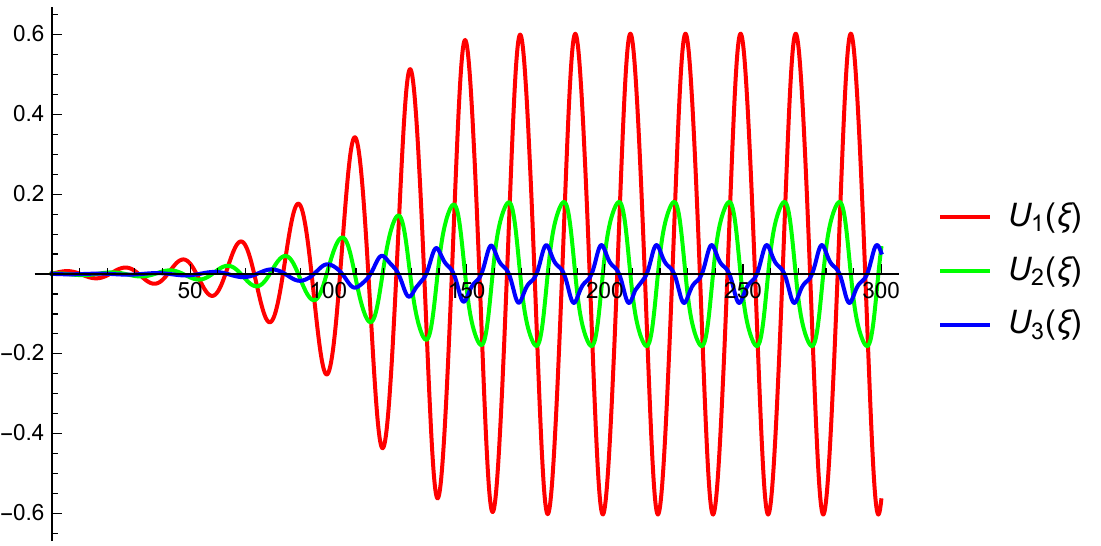}
     \includegraphics[width=0.475\textwidth]{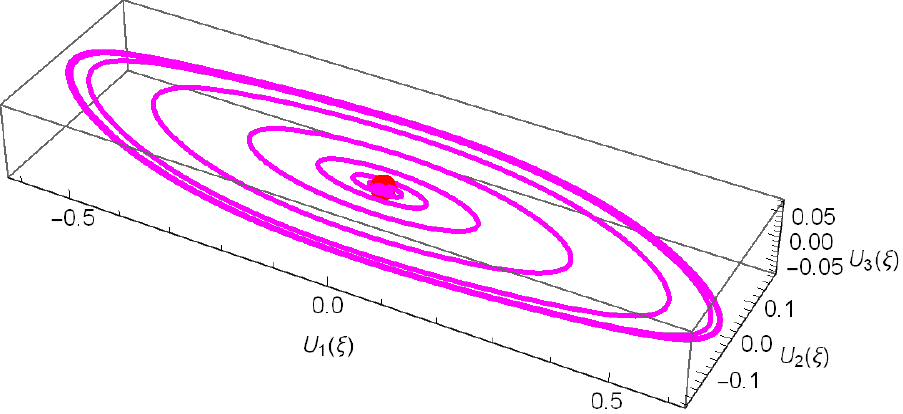}
    \caption{Case  C  for $n=2$. Periodic evolution  (left panel), stable periodic orbit (right panel).}
		\label{figD09}
\end{figure}

\end{document}